\documentstyle[prl,aps,floats,psfig]{revtex} 
\begin{document}
\twocolumn[
\hsize\textwidth\columnwidth\hsize\csname@twocolumnfalse\endcsname

\title{A proposal for a spin-polarized solar battery}

\author{Igor \v{Z}uti\'{c}$^1$, Jaroslav Fabian$^{1,2}$, and 
S. Das Sarma$^{1}$} 
\address{$^1$Department of Physics, University of Maryland at  College
Park, College Park, Maryland 20742-4111, USA\\
$^2$Max-Planck Institute for the Physics of Complex Systems, 
N\"{o}thnitzer Str. 38, D-01187 Dresden, Germany}

\maketitle

\begin{abstract}
A solar cell illuminated by circularly-polarized light generates
charge and spin currents. 
We show that the spin polarization of the  current
significantly exceeds the spin 
polarization of the carrier density for the
majority carriers. Based on this principle we propose a
semiconductor spin-polarized
solar battery and substantiate our proposal using analytical arguments
and numerical modeling.
\end{abstract}
\pacs{}
]

Illumination of a semiconductor sample 
by circularly-polarized light results in a spin polarization
of the carriers\cite{orientation84}.
Optical spin polarization of both minority (optical 
orientation) and majority (optical pumping) carriers
has been realized\cite{orientation84}. 
Introducing spin into semiconductors has also been
reported by injecting spin-polarized carriers from a 
magnetic material (metal\cite{hammar99} or semiconductor
\cite{fiederling99}). Combined with the existence of reasonably
long spin-relaxation times\cite{kikkawa98,fabian99}, this makes a strong
case for an all-semiconductor spintronics (traditional spintronic
devices are metallic\cite{prinz98}, suitable  
for their use in magnetic read heads and computer 
memory cells). The  
advantages of semiconductor spintronics would be
an easier integration with the existing semiconductor 
electronics and more versatile devices; 
for example, information storage and processing could, in
principle, be possible on the same spintronic chip.
There already exist theoretical proposals for semiconductor
unipolar spin transistors and spin diodes\cite{datta90,flatte01},
and bipolar semiconductor devices based on the spin-polarized 
{\it p-n} junction\cite{dassarma00,zutic01}.
Related experimental advances, 
demonstrating spin-polarized light-emitting 
diodes\cite{fiederling99} and a gate-voltage tunable magnetization in 
magnetic semiconductors\cite{ohno00a}, as well as the fabrication of 
a magnetic {\it p-n} junction 
\cite{ohno00}, provide further motivation to explore all-semiconductor
spintronics.

\begin{figure}
\centerline{\psfig{file=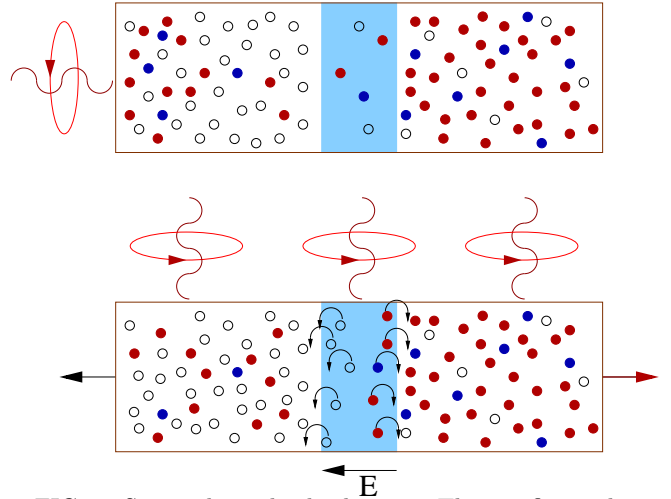,width=1\linewidth}}
\caption{Spin-polarized solar battery. The top figure shows a configuration
where a circularly-polarized light creates electron-hole pairs at the
$p$-region (left). Spin-polarized electrons 
[red (blue)=spin up (down)]
diffuse towards the depletion region where they are swept by the
built-in field $E$ to the $n$-side. Spin-polarization is pumped into
the majority region [10]. The bottom figure illustrates the 
geometry considered  in the text. A uniform illumination is assumed 
throughout the sample, giving rise to spin-polarized current 
(both spin and charge flow).
}
\label{fig:1}
\end{figure}

In this paper we propose a 
spin-polarized solar battery
as a source of both charge and spin currents. For its operation 
it is necessary to have spin imbalance in carrier population
(or in the corresponding components of current) 
as well as a
built-in field which separates  electron-hole pairs, created 
by illumination, producing voltage\cite{fahrenbruch83}. 
We consider a particular implementation of a spin-polarized
solar battery based on the concept
of spin-polarized {\it p-n} junction\cite{zutic01}.
A circularly-polarized light uniformly illuminates the sample 
(Fig.~\ref{fig:1}), generating spin-polarized carriers and  spin-polarized 
charge current. We reveal by numerical modeling
of drift-diffusion equations for spin and charge transport, 
that in the majority region current spin polarization is significantly 
enhanced over the carrier density polarization, 
and that spin polarization of the 
minority carriers 
near an ideal Ohmic contact is larger than 
in the bulk. By calculating the I-V 
characteristics for both charge and spin currents, we show that 
spin currents in the $n$-region generally diminish with increasing
forward voltage. We develop an analytical model based on spin diffusion
to further support these findings. 

\begin{figure}
\centerline{\psfig{file=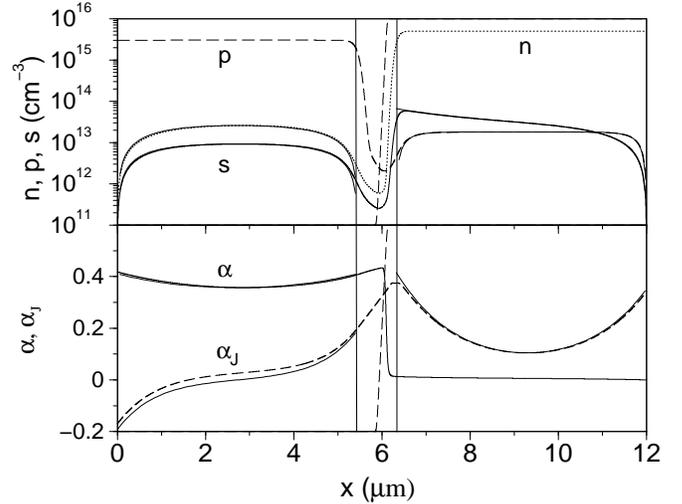,width=1\linewidth}}
\caption{Calculated spatial profiles of (top) carrier densities $n$ 
and $p$, spin density $s$, and (bottom) electron and 
current spin polarizations $\alpha$ and $\alpha_J$. The 
thin dashed lines show the doping profile $N_D(x)-N_A(x)$ (not to scale),
and the two vertical lines at $x_p\approx 5.4$ and $x_n\approx 6.3$
indicate the depletion layer boundaries.
The thin lines accompanying the numerical curves are analytical results 
for an ideal spin-polarized solar cell (if not visible, they overlap
with the numerical results).
}
\label{fig:2}
\end{figure}

Consider a GaAs sample at the room temperature, of length $L$ 
(extending on the $x$-axis from $x=0$ to $12$ $\mu$m), 
doped with $N_A=3\times 10^{15}$ cm$^{-3}$ acceptors on the left and with 
$N_D=5\times 10^{15}$ cm$^{-3}$ donors on the right [the doping profile,
$N_D(x)-N_A(x)$, is shown in Fig. \ref{fig:2}]. The intrinsic carrier 
concentration is $n_i=1.8\times 10^{6}$ cm$^{-3}$\cite{fahrenbruch83}, 
and the
electron (hole) mobility and diffusivity are $4000$ ($400$) cm$^2\cdot$ 
V$^{-1}\cdot$s$^{-1}$ and 103.6 (10.36) cm$^2\cdot$s$^{-1}$ 
\cite{fahrenbruch83}. The pair (band-to-band) recombination rate is taken 
to be $w=(1/3)\times 10^{-5}$ cm$^{3}\cdot$s$^{-1}$, giving an electron
lifetime in the $p$-region of $\tau_n=1/wN_A=0.1$ ns, and a hole lifetime
in the $n$-region of $\tau_p=1/wN_D=0.06$ ns. Spin relaxation time
(which is the spin lifetime in the $n$-region) is $T_1=0.2$ ns. 
In the $p$-region electron spin decays on the time scale of 
\cite{orientation84} $\tau_s=T_1\tau_n/(T_1+\tau_n)\approx 0.067$ ns.  
The minority diffusion lengths are $L_n=(D_n\tau_n)^{0.5}\approx 1$ $\mu$m
for electrons in the $p$-region, and $L_p=(D_p\tau_p)^{0.5}\approx 0.25$
$\mu$m for holes in the $n$-region. The spin decays on the length scale
of $L^{p}_s=(D_n\tau_s)^{0.5}\approx 0.8$ $\mu$m in the p- and 
$L^n_s=(D_nT_1)^{0.5} \approx 1.4$ $\mu$m in the n-region.
At no applied voltage, the depletion layer formed around $x_d=L/2=6$ 
$\mu$m has a width of $d\approx0.9$ $\mu$m, 
of it $d_p=(5/8)d$ in the $p$-side and
$d_n=(3/8)d$ in the $n$-side.

Let the sample be uniformly illuminated with a circularly-polarized light
with photon energy higher than the band gap (bipolar photogeneration).
The pair generation rate is chosen to be $G=3\times 10^{23}$ cm$^{-3}\cdot$
s$^{-1}$ (which corresponds to a concentrated solar light of intensity about 
1 W$\cdot$cm$^{-2}\cdot$s$^{-1}$),
so that in the bulk of the $p$-side there are $\Delta n=G\tau_n\approx 
3\times 10^{13}$ cm$^{-3}$ 
nonequilibrium electrons and holes; in the $n$-side the  density
is $\Delta p=G\tau_p=1.8\times 10^{13}$  cm$^{-3}$.
 Band structure of GaAs allows a 50\% spin polarization
of electrons excited by a circularly polarized light, so that the
spin-polarization at the moment of creation is $\alpha_0=G_s/G=0.5$, where
$G_s=G_\uparrow-G_\downarrow$ is the difference in the generation rates 
of spin up and down electrons. For a homogeneous doping, the spin
density in the $p$-side would be
$s_p=G_s\tau_s\approx 1\times 10^{13}$ cm$^{-3}$, while in 
the $n$-side $s_n=G_sT_1\approx 3\times 10^{13}$ cm$^{-3}$. 
Holes in GaAs can be considered unpolarized, since they lose
their spin on the time scale of momentum relaxation (typically a picosecond). 
The physical situation and the geometry are illustrated in Fig. \ref{fig:1},
bottom.

We solve numerically the drift-diffusion equations  
for inhomogeneously doped spin-polarized semiconductors\cite{zutic01}
to obtain electron and hole densities $n$ and $p$, spin density 
$s=n_\uparrow-n_\downarrow$ (where $n_\uparrow$ and $n_\downarrow$ are
spin up and down electron densities), and charge $J$ and spin $J_s=
J_\uparrow - J_\downarrow$ (where $J_\uparrow$ and $J_\downarrow$
are spin up and down electron charge currents) current densities. 
We consider ideal Ohmic contacts attached
at both ends of the sample, providing infinite carrier and spin recombination
velocities (so that both nonequilibrium carrier densities and spin density 
vanish at $x=0$ and $x=L$). 
Our sample is large enough 
(compared to
$L_n$, $L_p$, and $L_s$)
to distinguish the bulk from the 
boundary effects, so the behavior of more realistic boundary conditions
(which would include finite surface recombination velocities 
for both nonequilibrium carriers and spin) can be readily deduced
from our results.

\begin{figure}
\centerline{\psfig{file=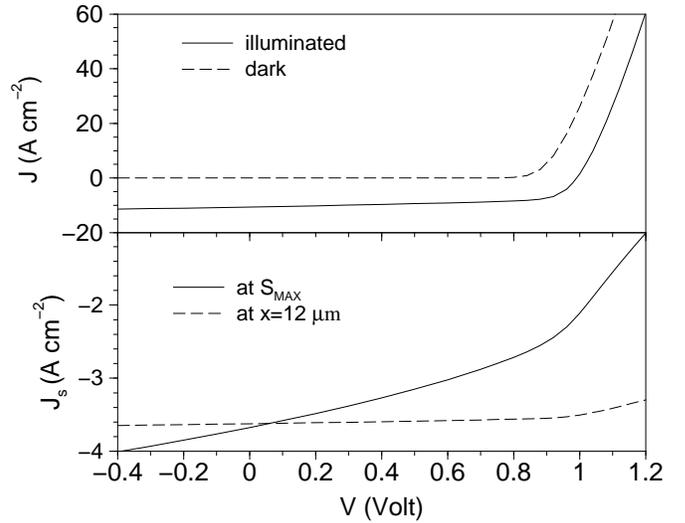,width=1\linewidth}}
\caption{Calculated I-V curves for charge currents (top: 
a solar cell in the dark and under illumination)
and spin currents (bottom: the solid curve is for
the spin current at the point in the $n$-region where $s$ is 
maximum--see Fig. \ref{fig:2}, and the dashed curve is for 
$J_s$ at the right end, $x=12$ $\mu$m).
}
\label{fig:3}
\end{figure}

Calculated spatial profiles of carrier and spin densities, as well
as carrier and current polarizations 
$\alpha=n/s$ and $\alpha_J=J_s/J$,  
are in Fig.~\ref{fig:2}. There is no applied voltage V, but the illumination
produces a reverse photo current 
$J_{\rm photo}=-eG(L_n+L_p+d)\approx -11$ A$\cdot$cm$^{-2}$ 
(see also Fig. \ref{fig:3}). The behavior of currier
densities is the same as in the unpolarized case (spin polarization
in nondegenerate semiconductors does not affect charge currents, as 
diffusivities for spin up and down carriers are equal). The spin density
essentially follows the nonequilibrium electronic density in the
$p$-side, sharply decreases in the depletion layer, while then rapidly
increasing to a value larger than the normal excitation value in the
$n$-side, $s_n$. We interpret
this as a result of spin pumping through the minority channel\cite{zutic01}:
electron spin excited within the distance $L^s_p$ from the depletion region, 
as well as generated inside that region, is swept into the $n$-side
by the built-in field, thus pumping spin polarization into the 
$n$-region. In the rest of the $n$-region, spin density 
decreases, until it reaches zero at the right boundary.
Carrier spin polarization $\alpha$ is reasonably high in the $p$-side,
but diminishes in the $n$-side.
[Note that in the geometry considered
in\cite{zutic01} (top of Fig. \ref{fig:1}), for a higher illumination
intensity and short  
junction, spin polarization remains almost unchanged
through the depletion layer, a result of a much more effective
electronic spin pumping.] Current polarization, however, remains quite
large throughout the sample. It changes sign in the $p$-region
(note that $\alpha_J=J_s/J$, and since $J(V=0)=J_{\rm photo}<0$  
is a constant, $\alpha_J$ shows the negative profile of spin current),
and has a symmetric shape in the $n$-region, being much larger than $\alpha$.

The profile of the carrier densities can be understood from the ideal solar
cell model, based on minority carrier diffusion, and Shockley
boundary conditions\cite{fahrenbruch83} 
(which, for $V=0$, states that the nonequilibrium
carrier density vanishes at the edges of the depletion layer). 
We do not write the formulas here, but we plot the analytical
results in Fig.~\ref{fig:2}. 
The behavior of $s(x)$
can be understood along
similar lines. Outside the depletion region we can neglect the
electric field as far as spin transport is considered (one does not
distinguish minority and majority spins--spin is everywhere out
of equilibrium, and it can be treated similarly to minority carrier 
densities). The equation for spin diffusion is
$D_n d^2 s/d x^2= (wp+1/T_1)s -G_s$.
Consider first the $p$-region. The boundary conditions are
$s(0)=0$ (the ideal Ohmic contact) and $s(x_p)=0$, where $x_p=x_d-d_p$ 
is the point where, roughly, the depletion layer begins 
(see Fig. \ref{fig:2}). 
The latter condition is an analogue of the Shockley condition 
that says that the photogenerated minority carrier density
vanishes at the edges of the depletion layer, as carriers generated there
are immediately swept into the other side of the layer by the built-in
field. The same reasoning holds for spin, as spin is carried
by the photogenerated electrons. The resulting spin density is
\begin{eqnarray}
s(x)=s_p\left [\frac{\cosh(\xi_p)-1}
{\sinh(\xi_p)}\sin(\xi)-\cosh(\xi)+1\right ],
\end{eqnarray}
where $\xi=x/L^p_s$ and $\xi_p=x_p/L^p_s$. The spin
current $J_s=\alpha_J J=eD_nds/dx$. These analytical results, 
plotted in Fig. \ref{fig:2}, agree with numerical calculation. 
Note that near %
the Ohmic contact spin polarization 
$\alpha(x\rightarrow 0)=\alpha_0 
(\tau_s/\tau_n)^{0.5}\approx 0.41$, which is
larger than the bulk value of $\alpha_0 (\tau_s/\tau_n)\approx 0.33$. 
The change in sign of $J_s$ is related to the increase of $s$ with 
increasing $x$, at small $x$,  
and then decrease close to the depletion layer. Current polarization is 
$\alpha_J(0)=-\alpha_0 L^p_s/(L_n+L_p+d)\approx -0.19$  
and $\alpha_J(x_p)= -\alpha_J(0)$. 

In the $n$-region, the right boundary value is that of an Ohmic 
contact, $s(L)=0$,
but at the left it is a finite value $s(x_n)=s_0$ (where $x_n$ is the
depletion region boundary with the $n$-side, $x_n=x_d+d_n$),
determined below. The solution of the diffusion equation 
is  
\begin{eqnarray}\label{eq:sn}
s(x)=s_n\left[\frac{\cosh(\eta_n)-1+s_0/s_n}
{\sinh(\eta_n)}\sinh(\eta)-\cosh(\eta)+1 \right ], 
\end{eqnarray}
where $\eta=(L-x)/L^n_s$ and $\eta_n=(L-x_n)/L^n_s$. To obtain
$s_0$, consider the physics which leads to its final value. In an
ideal case, all the electron spin generated in the 
$p$-region within the distance $L^p_s$ from the depletion
layer, as well as generated within the depletion layer, 
flow without relaxation into the $n$-region. Then 
the boundary condition for the spin current at $x_n$ reads
$J_s(x_n)=-eG_s(L^p_s+d)$. Since, at the same time, $J_s(x_n)=
eD_nds/dx|_{x_n}$, from Eq. \ref{eq:sn} we obtain 
$s_0=s_n[1+\tanh(\eta_n)(L^p_s+d)/L^n_s-1/\cosh(\eta_n)]$. 
In general, for a long junction ($\eta_n\gg 1)$, $s_0=
G_sT_1[1+(L^p_s+d)/L^n_s]$, and the enhancement of spin
due to the minority electron spin pumping is particularly
large for reverse biased samples with large $d$. For a short
junction ($\eta_n\ll 1$), $s_0=s_n\eta_n(L^p_s+d)/L^n_s$,
and the  spin at $x_n$ is solely due to electron spin 
pumping (but its value is smaller than for a long junction).
In our case $s_0\approx 2.2 s_n$, and $s(x)$ (Eq.~\ref{eq:sn}),
plotted in Fig. \ref{fig:2}, gives a very good agreement
with numerical data. Spin polarization of the current at 
the Ohmic contact is
$\alpha_J(L)=\alpha_0L^n_s/(L_n+L_p+d)\approx 0.33$, while
that at $x_n$ is $\alpha_J(x_n)=\alpha_0[(L^p_s+d)/(L_n+L_p+d)]
\approx 0.39$.
Current polarization is much larger than carrier polarization,
since both spin and charge currents are mainly diffusive. 
If only the $p$-region would be illuminated\cite{zutic01} with
photogenerated spin density $G_s$,
the induced spin density in the $n$-region would be
$s_0=G_s(T_1\tau_s)^{0.5}\tanh(\eta_n)$. This is purely the
minority-electron spin pumping effect. It is most effective
for long junctions, where the spin amplification is
$s_0/s_p=(T_1/\tau_s)^{0.5}=(1+T_1/\tau_n)^{0.5}$. At low
temperatures $T_1$ can be larger than $\tau_n$ by orders
of magnitude, and so spin amplification can be significant.

Finally, in Fig.~\ref{fig:3} we plot the I-V characteristics of the charge
and spin currents. The resulting charge I-V curve under illumination
can be, as in standard solar cells,  understood as the effect of 
superposition\cite{fahrenbruch83} of the negative short circuit current
(reverse photo current $J_{\rm photo}$) and the dark current, exponentially
increasing with forward voltage. The total charge current 
vanishes at the open-circuit voltage of about 1 V.
As spin current is {\em not} %
conserved (it varies in space), we choose two points to represent it on 
the I-V plot. One is the value of $J_s$ at the right boundary, the other
at the point where spin is maximum (at the right edge of the depletion
layer; this is an important point when a short junction would be considered).
Both values decrease in magnitude with increasing voltage, as a result of
decreasing of the effect of spin pumping from the nonequilibrium minority 
electrons. This is much more pronounced in the case of $J_s$ at maximum $s$,
which is most sensitive to the electronic pumping, as it varies with $d$ 
(which decreases with increasing voltage).

This work was supported by DARPA and the U.S. O.N.R.


\begin{references}

\bibitem{orientation84}  {\it Optical Orientation}, edited by
F. Meier and B. P. Zakharchenya (North-Holland, New York 1984).

\bibitem{hammar99} P. R. Hammar, B. R. Bennet, M. J. Yang, and M. Johnson,
{Phys. Rev. Lett.} {\bf 83}, 203 (1999).

\bibitem{fiederling99}
R. Fiederling, M. Kleim, G. Reuscher, W. Ossau, G. Schmidt, A. Waag,
and L. W. Molenkamp, {\sl Nature} {\bf 402}, 787 (1999);
Y. Ohno, D. K. Young, B. Beschoten, F. Matsukura, H. Ohno, and D. D.
Awschalom, {\sl Nature} {\bf 402}, 790 (1999);
B. T. Jonker Y. D. Park, B. R. Bennett, H. D. Cheong,
G. Kioseoglou, and A. Petrou, Phys. Rev. B {\bf 62}, 8180 (2000).

\bibitem{kikkawa98}
J. M. Kikkawa and D. D. Awschalom, Phys. Rev. Lett. {\bf 80}, 4313 (1998).

\bibitem{fabian99}
J. Fabian and S. Das Sarma, J. Vac. Sc. Technol. B {\bf 17}, 1708 (1999).

\bibitem{prinz98} G. A. Prinz, {\sl Science} {\bf 282},
1660 (1998).

\bibitem{datta90} S. Datta and B. Das, 
Appl. Phys. Lett. {\bf 56}, 665 (1990).

\bibitem{flatte01} M. E. Flatte and G. Vignale, Appl. Phys. Lett. {\bf 78},
1273 (2001).
 
\bibitem{dassarma00} S. Das Sarma, J. Fabian, X. D. Hu, and  I. \v{Z}uti\'{c}, 
58th Device Research Conference, p. 95 (IEEE; Piscataway 2000) and LANL 
Preprint cond-mat/0006369.

\bibitem{zutic01} I. \v{Z}uti\'{c}, J. Fabian, and S. Das Sarma, 
LANL Preprint cond-mat/0104146.

\bibitem{ohno00a} H. Ohno, D. Chiba, F. Matsukura, T. Omiya, E. Abe,
T. Dietl, Y. Ohno, and K. Ohtani, {\sl Nature} {\bf 408}, 944 (2000).

\bibitem{ohno00} Y. Ohno, L. Arata, F. Matsukura, S. Wang, and H. Ohno,
Appl. Surf. Science {\bf 159}, 308 (2000).

\bibitem{fahrenbruch83} A. L. Fahrenbruch and R. H. Bube, {\it
Fundamentals of Solar Cells} (Academic, 1983).

\end{references}
\end{document}